\documentclass[a4paper,oneside, onecolumn,final,times,number,3p]{elsarticle}
\usepackage{amsmath,amsthm}
\usepackage{mathtools}
\usepackage{framed}
\usepackage{multicol} 
\usepackage{nomencl} 

\makenomenclature

\renewcommand*\nompreamble{\begin{multicols}{2}}

\renewcommand*\nompostamble{\end{multicols}}
\usepackage{graphics}
\usepackage{epsfig}
\usepackage{float}
\usepackage{epstopdf}
\usepackage{enumitem}
\graphicspath{{./images/}}

\usepackage{graphicx}
\usepackage{caption}
\usepackage{subcaption}

\makeatletter
\g@addto@macro\@floatboxreset\centering
\makeatother
\newcommand{\Ree}{\operatorname{Re}}
\makeatletter
\@fpsep\textheight
\makeatother
\extrafloats{100}
\usepackage{caption}
\journal{   }

\begin{document}
\begin{frontmatter}

\title {Numerical solutions of an unsteady 2-d incompressible flow with heat and mass transfer at low, moderate, and high Reynolds numbers}

\author[rvt]{V. Ambethkar\corref{cor1}}
\ead{vambethkar@maths.du.ac.in, vambethkar@gmail.com}
\author[rvt]{Durgesh Kushawaha}
\ead{durgeshoct@gmail.com}

\cortext[cor1]{Corresponding author}

\address[rvt]{Department of Mathematics,\\
Faculty of Mathematical Sciences,\\
University of Delhi\\ Delhi,
110007, India}

\begin{abstract}

In this paper, we have proposed a modified Marker-And-Cell (MAC) method to investigate the problem of an unsteady 2-D incompressible flow with heat and mass transfer at low, moderate, and high Reynolds numbers with no-slip and slip boundary conditions. We have used this method to solve the governing equations along with the boundary conditions and thereby to compute the flow variables, viz. $u$-velocity, $v$-velocity, $P$, $T$, and $C$. We have used the staggered grid approach of this method to discretize the governing equations of the problem. A modified MAC algorithm was proposed and used to compute the numerical solutions of the flow variables for Reynolds numbers $Re=10$, $500$, and $50,000$ in consonance with low, moderate, and high Reynolds numbers. We have also used appropriate Prandtl ($Pr$) and Schmidt ($Sc$) numbers in consistence with relevancy of the physical problem considered. We have executed this modified MAC algorithm with the aid of a computer program developed and run in C compiler. We have also computed numerical solutions of local Nusselt ($Nu$) and Sherwood ($Sh$) numbers along the horizontal line through the geometric center at low, moderate, and high Reynolds numbers for fixed $Pr=6.62$ and $Sc=340$ for two grid systems at time $t=0.0001s$. Our numerical solutions for $u$ and $v$ velocities along the vertical and horizontal line through the geometric center of the square cavity for $Re=100$ has been compared with benchmark solutions available in the literature and it has been found that they are in good agreement. The present numerical results indicate that, as we move along the horizontal line through the geometric center of the domain, we observed that, the heat and mass transfer decreases up to the geometric center. It, then, increases symmetrically.
\end{abstract}

\end{frontmatter}

\section{Introduction}
\label{sec1}

The problem of 2-D unsteady incompressible viscous fluid flow with heat and mass transfer has been the subject of intensive numerical computations in recent years. This is due to its significant applications in many scientific and engineering practices. Fluid flows play an important role in various equipment and processes. Unsteady 2-D incompressible viscous flow coupled with heat and mass transfer is a complex problem of great practical significance. This problem has received considerable attention due to its numerous engineering practices in various disciplines, such as storage of radioactive nuclear waste materials, transfer groundwater pollution, oil recovery processes, food processing, and the dispersion of chemical contaminants in various processes in the chemical industry. More often, fluid flow with heat and mass transfer are coupled in nature. Heat transfer is concerned with the physical process underlying the transport of thermal energy due to a temperature difference or gradient. All the process equipment used in engineering practice has to pass through an unsteady state in the beginning when the process is started, and, they reach a steady state after sufficient time has elapsed. Typical examples of unsteady heat transfer occur in heat exchangers, boiler tubes, cooling of cylinder heads in I.C. engines, heat treatment of engineering components and quenching of ingots, heating of electric irons, heating and cooling of buildings, freezing of foods, etc. Mass transfer is an important topic with vast industrial applications in mechanical, chemical and aerospace engineering. Few of the applications involving mass transfer are absorption and desorption, solvent extraction, evaporation of petrol in internal combustion engines etc. Numerous everyday applications such as dissolving of sugar in tea, drying of wood or clothes, evaporation of water vapor into the dry air, diffusion of smoke from a chimney into the atmosphere, etc. are also examples of mass diffusion. In many cases, it is interesting to note that heat and mass transfer occur simultaneously.

Harlow and Welch \cite{b1} used the Marker-And-cell (MAC) method for numerical calculation of time-dependent viscous incompressible flow of fluid with the free surface. This method employs the primitive variables of pressure and velocity that has practical application to the modeling of fluid flows with free surfaces. Ghia et al. \cite{b3} have used the vorticity-stream function formulation for the two-dimensional incompressible Naiver-Stokes equations to study the effectiveness of the coupled strongly implicit multi-grid (CSI-MG) method in the determination of high-Re fine-mesh flow solutions. Issa et al. \cite{b5} have used a non-iterative implicit scheme of finite volume method to study compressible and incompressible recirculating flows. Elbashbeshy \cite{b6} has investigated the unsteady mass transfer from a wedge. Sattar \cite{b7} has studied free convection and mass transfer flow through a porous medium past an infinite vertical porous plate with time-dependent temperature and concentration. Sattar and Alam \cite{b8} have investigated the MHD free convective heat and mass transfer flow with hall current and constant heat flux through a porous medium. Maksym Grzywinski, Andrzej Sluzalec \cite{b10} solved the stochastic convective heat transfer equations in finite differences method. A numerical procedure based on the stochastic finite differences method was developed for the analysis of general problems in free/forced convection heat transfer. Lee \cite{b11} has studied fully developed laminar natural convection heat and mass transfer in a partially heated vertical pipe. Chiriac and Ortega \cite{b12} have numerically studied the unsteady flow and heat transfer in a transitional confined slot jet impinging on an isothermal plate. De and Dalai \cite{b14} have numerically studied natural convection around a tilted heated square cylinder kept in an enclosure has been studied in the range of $\rm 1000\le Ra\le 1000000$. Detailed flow and heat transfer features for two different thermal boundary conditions are reported. Chiu et al.
 \cite{b15} proposed an effective explicit pressure gradient scheme implemented in the two-level non-staggered grids for incompressible Navier-Stokes equations. Lambert et al. \cite{b17} studied the heat transfer enhancement in oscillatory flows of Newtonian and viscoelastic fluids. Alharbi et al. \cite{b18} presented the study of convective heat and mass transfer characteristics of an incompressible MHD visco-elastic fluid flow immersed in a porous medium over a stretching sheet with chemical reaction and thermal stratification effects. Xu et al. \cite{b20} have investigated the unsteady flow with heat transfer adjacent to the finned sidewall of a differentially heated cavity with conducting adiabatic fin. Fang et al. \cite{b21} have investigated the steady momentum and heat transfer of a viscous fluid flow over a stretching/shrinking sheet. Salman et al. \cite{b22} have investigated heat transfer enhancement of nano fluids flow in micro constant heat flux. Hasanuzzaman et al. \cite{b23} investigated the effects of Lewis number on heat and mass transfer in a triangular cavity. Schladow \cite{b24} has investigated oscillatory motion in a side-heated cavity. Lei and Patterson \cite{b25} have investigated unsteady natural convection in a triangular enclosure induced by absorption of radiation. Ren and Wan \cite{b26} have proposed a new approach to the analysis of heat and mass transfer characteristics for laminar air flow inside vertical plate channels with falling water film evaporation.

The above mentioned literature survey pertinent to the present problem under consideration revealed that, to obtain high accurate numerical solutions of the flow variables, we need to depend on high accurate and high resolution method like the  modified Marker-And-Cell (MAC) method, being proposed in this work. Furthermore, a modified MAC algorithm is employed for computing unknown variables $u$, $v$, $P$, $T$, and $C$ simultaneously.

What motivated us is the enormous scope of applications of unsteady incompressible flow with heat and mass transfer as discussed earlier. Literature survey also revealed that, the problem of 2-D unsteady incompressible flow with, heat and mass transfer in a rectangular domain, along with slip wall, temperature, and concentration boundary conditions has not been studied numerically. Furthermore, it has also been observed that, there is no literature to conclude availability of high accurate method that solves the governing equations of the present problem subject to the initial and boundary conditions. Moreover, in order to investigate the importance of the applications enumerated upon earlier, there is a need to determine numerical solutions of the unknown flow variables. In order to fulfill this requirement, we present numerically an investigation of the problem of unsteady 2-D incompressible flow, with heat and mass transfer in a rectangular domain, along with slip wall, temperature, and concentration boundary conditions, using the modified Marker-And-Cell (MAC) method. Though there is a well known MAC method for solving the problem of 2-D fluid flow, we present a suitably modified MAC method as well as a modified MAC algorithm to compute the accurate numerical solutions of the flow variables to the problem considered in this work.

Our main target of this work is to propose and use the modified Marker-And-Cell (MAC) method of pressure correction approach to investigate the problem of unsteady 2-D incompressible flow with heat and mass transfer. We have proposed and used this method to solve the governing equations along with no-slip and slip wall boundary conditions and thereby to compute the flow variables. We have used a modified MAC algorithm for discretizing the governing equations in order to compute the numerical solutions of the flow variables at different Reynolds numbers in consonance with low, moderate, and high. We have executed this modified MAC algorithm with the aid of a computer program developed and run in C compiler. We have also computed numerical solutions of local Nusselt $(Nu)$ and Sherwood $(Sh)$ numbers along the horizontal line through the geometric center at low, moderate, and high $Re$, for fixed $Pr=6.62$ and $Sc=340$ for two grid systems at time $t=0.0001s$.

The summary of the layout of the current work is as follows: Section 2 describes mathematical formulation that includes physical description of the problem, governing equations, and initial and boundary conditions. Section 3 describes the modified Marker-And-Cell method, along with the discretization of the governing equations. Section 4 describes a modified Marker-And-Cell (MAC) algorithm, along with numerical computations. Section 5 discusses the numerical results. Section 6 illustrates the conclusions of this study. Section 7 provides validity of our computer code used to obtain numerical solutions with the benchmark solutions.

\section{Mathematical Formulation}
\label{sec2}
\subsection{Physical Description}
\label{sec2.1.}

\begin{figure}[htb]
\includegraphics[scale=1.2]{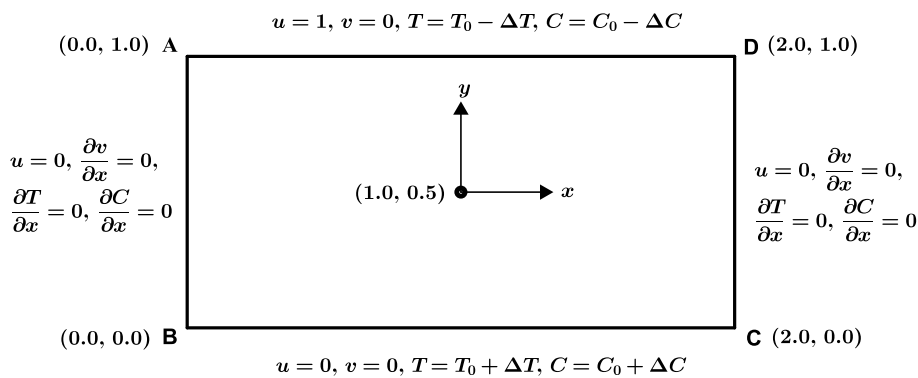}
\caption{Rectangular cavity}
\label{f1}
\end{figure}

Figure~\ref{f1} illustrates the geometry of the problem considered in this study along with no-slip and slip boundary conditions. ABCD is a rectangular domain around the point $(1.0,0.5)$ in which an unsteady 2-D incompressible viscous flow with heat and mass transfer is considered. Flow is setup in a rectangular domain with three stationary walls and a top lid that moves to the right with constant speed $(u=1)$.

We have assumed that, at all four corner points of the computational domain, velocity components $(u,v)$ vanish. It may be noted here regarding specifying the boundary conditions for pressure, the convention followed is that either the pressure at boundary is given or velocity components normal to the boundary is specified \cite[pp.129]{b2}.

\subsection{Governing equations}
\label{sec2.2.}

The governing equations of 2-D unsteady incompressible flow with heat and mass transfer in a rectangular domain are the continuity equation, the two components of momentum  equation, the energy equation, and the equation of mass transfer. These equations \eqref{e1} to \eqref{e5} subject to boundary conditions \eqref{e6} and \eqref{e7} are discretized using the modified Marker-And-Cell (MAC) method. Taking usual the Boussinesq approximations into account, the dimensionless governing equations are expressed as follows:
\begin{align}
&\mbox{Continuity equation}
&&
\dfrac{\partial u}{\partial x}+\dfrac{\partial v}{\partial y}=0,
\label{e1}
\\[2mm]
&\mbox{$x$-momentum}
&&
\dfrac{\partial u}{\partial t} +u\dfrac{\partial u}{\partial x}+v\dfrac{\partial u}{\partial y}=-
\dfrac{\partial P}{\partial x}+
\left(\dfrac{1}{Re}\right) \left(
\dfrac{\partial^2 u}{\partial x^2}+
\dfrac{\partial^2 u}{\partial y^2}\right),
\label{e2}
\\[2mm]
&\mbox{$y$-momentum}
&&
\dfrac{\partial v}{\partial t}+
u\dfrac{\partial v}{\partial x}+
v\dfrac{\partial v}{\partial y}=-
\dfrac{\partial P}{\partial y}+
\left(\dfrac{1}{Re}\right) \left(
\dfrac{\partial^2 v}{\partial x^2}+
\dfrac{\partial^2 v}{\partial y^2}\right),
\label{e3}
\\[2mm]
&\mbox{Energy equation}
&&
\dfrac{\partial T}{\partial t}+
u\dfrac{\partial T}{\partial x}+
v\dfrac{\partial T}{\partial y}=\left(\dfrac{1}{Pr}\right)
\left(\dfrac{\partial^2 T}{\partial x^2}+
\dfrac{\partial^2 T}{\partial y^2}\right),
\label{e4}
\\[2mm]
&\mbox{Mass transfer equation}
&&
\dfrac{\partial C}{\partial t}+
u\dfrac{\partial C}{\partial x}+
v\dfrac{\partial C}{\partial y}=\left(\dfrac{1}{Sc}\right)
\left(\dfrac{\partial^2 C}{\partial x^2}+
\dfrac{\partial^2 C}{\partial y^2}\right).
\label{e5}
\end{align}
where $u$, $v$, $P$, $T$, $C$, $Re$, $Pr$, and, $Sc$ are the velocity components in $x$ and $y$- directions, the pressure, the temperature, the concentration, the Reynolds number, the Prandtl number, and the Schmidt number respectively.\medskip

The initial, no-slip and slip wall boundary conditions are given by:
\begin{align}
&\text{for $t= 0$,}~~
u(x,y,0)=0, \ v(x,y,0)=0, \  T(x,y,0)=10, \  C(x,y,0)=10.
\label{e6}
\\[2mm]
&\!\!\!\!\left.
\begin{array}{lll}
\text{for $t>0$,}&
\text{on boundary AB: $u=0$, $\dfrac{\partial v}{\partial x}=0$,  $\dfrac{\partial T}{\partial x}=0$,  $\dfrac{\partial C}{\partial x}=0$,}
\\[4mm]
&\text{on boundary BC: $u=0$, $v=0$, $T=T_{0}+ \Delta{T}$, $C=C_{0}+ \Delta{C}$,}
\\[2mm]
&\text{on boundary CD: $u=0$, $\dfrac{\partial v}{\partial x}=0$,  $\dfrac{\partial T}{\partial x}=0$,  $\dfrac{\partial C}{\partial x}=0$,}
\\[4mm]
&\text{on boundary AD: $u=1$, $v=0$, $T=T_{0}- \Delta{T}$, $C=C_{0}- \Delta{C}$.}
\end{array}\right\}
\label{e7}
\end{align}

\section{Numerical Method and Discretization}
\label{sec3.}
\subsection{Modified Marker-And-Cell (MAC) Method}
\label{sec31}

\begin{figure}[htb]
\includegraphics[scale=0.85]{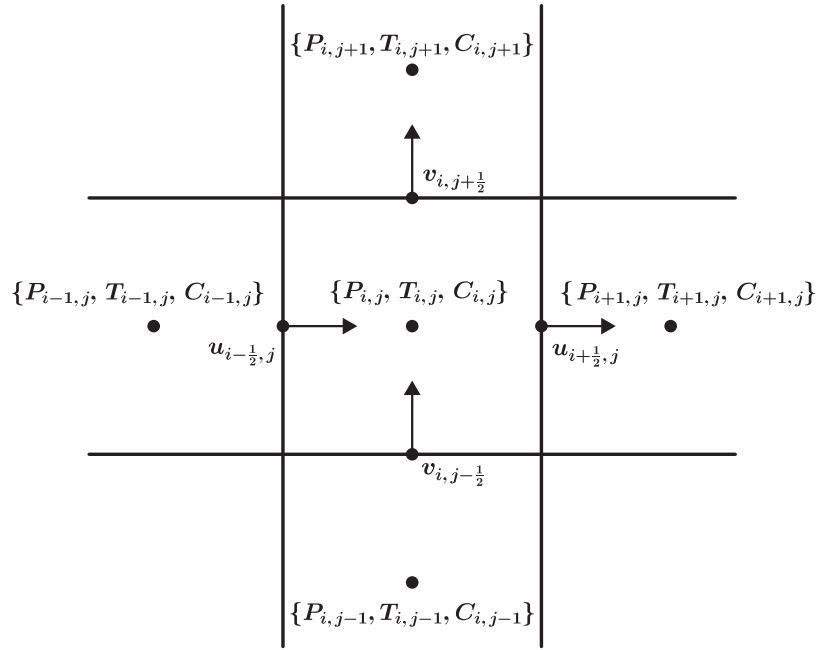}
\caption{Modified MAC staggered grid system}
\label{f2}
\end{figure}

Our main purpose in this work is to propose and use an accurate numerical method that solves the governing equations of the present problem subject to the initial and boundary conditions. In order to solve the equations \eqref{e1}-\eqref{e5} which are semi-linear coupled partial differential equations, we propose and use the modified Marker-And-Cell (MAC) method based on the MAC method of Harlow and Welch \cite{b1}. Consider a modified MAC staggered grid for $u$, $v$, and a scalar node ($P$ node), where pressure, temperature, and concentration variables are stored as shown in Figure \ref{f2}. The $x$-momentum equation is written at $u$ nodes, and the $y$-momentum equation is written at $v$ nodes.The energy and the mass transfer equations are written at a scalar node. Accordingly, the various derivatives in the $x$-momentum equation \eqref{e2} are calculated as follows:
\begin{align}
&\left(\dfrac{\partial u}{\partial t} \right)^{n+1}_{i+1/2,j}=
\Big(u_{i+1/2,j}^{n+1}-
u_{i+1/2,j}^{n}\Big)\Big/\Delta t,
\label{e8}
\\[1mm]
&\left(\dfrac{\partial^2 u}{\partial x^2} \right)^{n+1}_{i+1/2,j}=
\Big(u_{i+3/2,j}^{n+1}-2u_{i+1/2,j}^{n+1}+
u_{i-1/2,j}^{n+1} \Big)\Big/{\Delta x^2},
\label{e9}
\\[1mm]
&\left(\dfrac{\partial^2 u}{\partial y^2} \right)^{n+1}_{i+1/2,j}= \Big(u_{i+1/2,j+1}^{n+1}-2u_{i+1/2,j}^{n+1}+
u_{i+1/2,j-1}^{n+1} \Big)\Big/{\Delta y^2},
\label{e10}
\\[1mm]
&\left(\dfrac{\partial P}{\partial x} \right)^{n+1}_{i+1/2,j}=
\Big(P_{i+1,j}^{n+1}-
P_{i,j}^{n+1}\Big)\Big/\Delta x.
\label{e11}
\end{align}

Using the modified MAC method, the other terms in the $x$-momentum equation \eqref{e2} are written as
\begin{align}
&\left(u\dfrac{\partial u}{\partial x} \right)^{n}_{i+1/2,j}=
 u_{i+1/2,j}^{n}
 \Big(u_{i+3/2,j}^{n}-
 u_{i+1/2,j}^{n}\Big)\Big/\Delta x,
 \label{e12}
 \\[1mm]
 &\left(v\dfrac{\partial u}{\partial y} \right)^{n}_{i+1/2,j}=
 v_{i+1/2,j}^{n}
 \Big(u_{i+1/2,j+1}^{n}-
 u_{i+1/2,j}^{n}\Big)\Big/\Delta y.
 \label{e13}
 \end{align}
Here $v_{i+1/2,j}^n$ is given by
\begin{gather}
v_{i+1/2,j}^n= \Big(v_{i,j+1/2}^n+ v_{i,j-1/2}^n+v_{i+1,j+1/2}^n+
v_{i+1,j-1/2}\Big)\Big/4.
\label{e14}
\end{gather}

For simplicity, we will implement a fully explicit version of the time-splitting (fractional time-step) method, for both the viscous and the diffusion terms. When this method is applied to the $x$-momentum equation on the staggered grid from time level $t^n$ to $\hat{t}$ yields the following equation at the intermediate step:
\begin{gather}
\begin{gathered} \dfrac{\hat{u}_{i+1/2,j}-u_{i+1/2,j}^n}{\Delta t}= u_{i+1/2,j}^n
\dfrac{u^n_{i+1/2,j}-u_{i+3/2,j}^n}{\Delta x}+ v_{i+1/2,j}^n
\dfrac{\big(u^n_{i+1/2,j}-
u_{i+1/2,j+1}^n\big)}{\Delta y}
\\[1mm]
+ \dfrac{u_{i+3/2,j}^n-2u_{i+1/2,j}^n+
u_{i-1/2,j}^n}{R_e\Delta x^2}+
\dfrac{u_{i+1/2,j+1}^n-2u_{i+1/2,j}^n+
u_{i+1/2,j-1}^n}{R_e\Delta y^2}.
\end{gathered}
\label{e15}
\end{gather}
Similarly for the $y$-momentum equation we obtain the following equation:
 \begin{gather}
\begin{gathered} \dfrac{\hat{v}_{i,j+1/2}-v_{i,j+1/2}^n}{\Delta t}= u_{i,j+1/2}^n \dfrac{v^n_{i-1,j+1/2}-v_{i,j+1/2}^n}{\Delta x}+v_{i,j+1/2}^n
\dfrac{v^n_{i,j+1/2}-v_{i,j+3/2}^n}{\Delta y}
\\[1mm]
+ \dfrac{v_{i+1,j+1/2}^n-2v_{i,j+1/2}^n+
  v_{i-1,j+1/2}^n}{R_e\Delta x^2}+
 \dfrac{v_{i,j+3/2}^n-2v_{i,j+1/2}^n+
  v_{i,j-1/2}^n}{R_e\Delta y^2}.
\end{gathered}
\label{e16}
\end{gather}
Practical stability requirement obtained from the Von Neumann analysis for the Euler explicit solvers are given by Peyret and Taylor \cite[148]{b4}, as follows:
\begin{align}
&\big(|u|+|v|\big)^2\Delta tRe\le 4,
\label{e17}
\\[1mm]
&\dfrac{\Delta t}{\Ree}\left[
\dfrac{1}{\Delta x^2}+\dfrac{1}{\Delta y^2} \right]\le 5,
\label{e18}
\\[1mm]
&\Delta t_{i,j}^{\max}
\left[\left(\dfrac{u_{i,j}}{\Delta x}+
\dfrac{v_{i,j}}{\Delta y} \right)
\dfrac{1}{R_e}+\dfrac{2}{Pr} \left(
\dfrac{1}{\Delta x^2}+
\dfrac{1}{\Delta y^2}
\right) \right]\le 1,
\label{e19}
\\[1mm]
&\Delta t_{i,j}^{\max}
\left[\left(\dfrac{u_{i,j}}{\Delta x}+
\dfrac{v_{i,j}}{\Delta y} \right)
\dfrac{1}{R_e}+\dfrac{2}{Sc} \left(
\dfrac{1}{\Delta x^2}+
\dfrac{1}{\Delta y^2}
\right) \right]\le 1.
\label{e20}
\end{align}
Now advancing from $t^n$ to $\hat{t}$ and then $\hat{t}$ to $t^{n+1}$ one obtains the elliptical pressure equation:
\begin{gather}
\dfrac{\nabla \cdot \hat{u}}{\Delta t}=
\nabla^2 P^{n+1}
\label{e21}.
\end{gather}
The corresponding homogeneous Neumann boundary condition for pressure is given by:
\begin{gather}
\dfrac{\partial p^{n+1}}{\partial n}=0
\label{e22}.
\end{gather}

Now using central difference scheme:
\begin{gather}
\begin{multlined}
\dfrac{P^{n+1}_{i+1,j}-2P^{n+1}_{i,j}+
P^{n+1}_{i-1,j}}{\Delta x^2}+
\dfrac{P^{n+1}_{i,j+1}-2P^{n+1}_{i,j}+
P^{n+1}_{i,j-1}}{\Delta y^2}=
\\[1mm]
\qquad\qquad \qquad=\dfrac{1}{\Delta t}
\left[ \dfrac{\hat{u}_{i+1/2,j}-
\hat{u}_{i-1/2,j}}{\Delta x}+
\dfrac{\hat{v}_{i,j+1/2}-
\hat{v}_{i,j-1/2}}{\Delta y}\right].
\label{e23}
\end{multlined}
\end{gather}
We obtain the velocity field at the advanced time level $(n+1)$, for each velocity component, this equation gives
\begin{gather}
u_{i+1/2,j}^{n+1}=
\hat{u}_{i+1/2,j}-
\dfrac{\Delta t}{\Delta x}
\big(P_{i+1,j}^{n+1}-
P_{i,j}^{n+1}\big),
\label{e24}
\\[1mm]
v_{i,j+1/2}^{n+1}=
\hat{v}_{i,j+1/2}-
\dfrac{\Delta t}{\Delta y}
\big(P_{i,j+1}^{n+1}-
P_{i,j}^{n+1}\big).
\label{e25}
\end{gather}
The modified MAC staggered grid quotients of different derivatives which appeared in the energy equation is written as follows:
\begin{align}
&\left(\dfrac{\partial T}{\partial t} \right)^{n+1}_{i,j}=
\big(T_{i,j}^{n+1}-
T_{i,j}^{n}\big)/\Delta t,
\notag
\\[1mm]
&\left(\dfrac{\partial T}{\partial x} \right)^{n}_{i+1,j}=
\big(T_{i+1,j}^{n}-
T_{i,j}^{n}\big)/\Delta x,
\notag
\\[1mm]
&\left(\dfrac{\partial T}{\partial y} \right)^{n}_{i,j+1}=
\big(T_{i,j+1}^{n}-
T_{i,j}^{n}\big)/\Delta y,
\notag
\\[1mm]
&\left(\dfrac{\partial^2 T}{\partial x^2} \right)^{n}_{i+1,j}=
\big(T_{i+1,j}^{n}-2T_{i,j}^{n}+
T_{i-1,j}^{n}\big)/ \Delta x^2,
\notag
\\[1mm]
&\left(\dfrac{\partial^2 T}{\partial y^2} \right)^{n}_{i,j+1}=
\big(T_{i,j+1}^{n}-2T_{i,j}^{n}+
T_{i,j-1}^{n}\big)/ \Delta y^2.
\notag
\end{align}
The discretized  form of the energy equation \eqref{e4} is given by:
\begin{gather}
\begin{multlined} \dfrac{T^{n+1}_{i,j}-T^{n}_{i,j}}{\Delta t}=u_{i,j}^n \dfrac{T^{n}_{i,j}-T^{n}_{i+1,j}}{\Delta x}+v_{i,j}^n \dfrac{T^{n}_{i,j}-T^n_{i,j+1}}{\Delta y}
\\[1mm]
\hspace{35mm}+\dfrac{T_{i+1,j}^n-2T_{i,j}^n+
T^n_{i-1,j}}{Pr \Delta x^2}+
\dfrac{T_{i,j+1}^n-2T_{i,j}^n+
T^n_{i,j-1}}{Pr \Delta y^2}.
\end{multlined}
\label{e26}
\end{gather}
The modified MAC staggered grid quotients of different derivatives which appeared in the mass transfer equation is written as follows:
\begin{align*}
&\left(\dfrac{\partial C}{\partial t} \right)^{n+1}_{i,j}=
\big(C_{i,j}^{n+1}-
C_{i,j}^{n}\big)/\Delta t,
\notag
\\[1mm]
&\left(\dfrac{\partial C}{\partial x} \right)^{n}_{i+1,j}=
\big(C_{i+1,j}^{n}-
C_{i,j}^{n}\big)/\Delta x,
\notag
\\[1mm]
&\left(\dfrac{\partial C}{\partial y} \right)^{n}_{i,j+1}=
\big(C_{i,j+1}^{n}-
C_{i,j}^{n}\big)/\Delta y,
\\[1mm]
&\left(\dfrac{\partial^2 C}{\partial x^2} \right)^{n}_{i+1,j}=
\big(C_{i+1,j}^{n}-2C_{i,j}^{n}+
C_{i-1,j}^{n}\big)/ \Delta x^2,
\\[1mm]
&\left(\dfrac{\partial^2 C}{\partial y^2} \right)^{n}_{i,j+1}=
\big(C_{i,j+1}^{n}-2C_{i,j}^{n}+
C_{i,j-1}^{n}\big)/ \Delta y^2.
\end{align*}

The discretized  form of the mass transfer equation \eqref{e5} is given by:
\begin{gather}
\begin{multlined}
\dfrac{C^{n+1}_{i,j}-C^{n}_{i,j}}{\Delta t}=u_{i,j}^n
\dfrac{C^{n}_{i,j}-C^{n}_{i+1,j}}{\Delta x}+v_{i,j}^n \dfrac{C^{n}_{i,j}-C^n_{i,j+1}}{\Delta y}
\\[1mm]
\hspace{33mm}+\dfrac{C_{i+1,j}^n-2C_{i,j}^n+
  C^n_{i-1,j}}{Sc \Delta x^2}+
 \dfrac{C_{i,j+1}^n-2C_{i,j}^n+
  C^n_{i,j-1}}{Sc \Delta y^2}
\end{multlined}
\label{e27}
\end{gather}
We note that $u_{i,j}^n$ is not defined on $u$ node and $v_{i,j}^n$ is not defined on $v$ node. Therefore in order to obtain these quantity, we use averaging as given below:
\begin{gather}
u_{i,j}^n=\dfrac{1}{2}
\Big(u_{i+1/2,j}^n+
u_{i-1/2,j}^n\Big)
\quad\mbox{and}\quad
v_{i,j}^n=\dfrac{1}{2}
\Big(v_{i,j+1/2}^n+
v_{i,j-1/2}^n\Big).
\label{e28}
\end{gather}

\section{Numerical Computations}
\label{sec4.}

We have used a modified MAC algorithm to the discretized governing equations in order to compute the numerical solutions of the flow variables at different Reynolds numbers in consonance with low, moderate, and high. We have executed this modified MAC algorithm with the aid of a computer program developed and run in C compiler. The input data for the relevant parameters in the governing equations like Reynolds number ($Re$), Prandtl Number ($Pr$), and Schmidt number ($Sc$) has been properly chosen incompatible with the present problem considered.

\subsection{Modified MAC Algorithm}
\label{sec4.1}

We summarize the sequence of computational steps involved in the modified MAC algorithm as follow:

\begin{figure}[htb]
\includegraphics[scale=0.65]{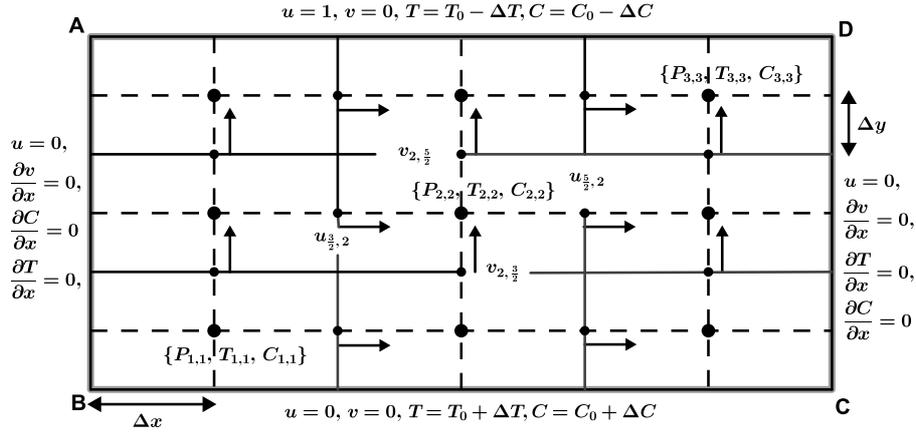}
\caption{The rectangular staggered computational grid}
\label{f3}
\end{figure}

\subsubsection{Prediction Step}
\bigskip
\begin{itemize}
\item Using \eqref{e15} and \eqref{e16}, calculate $\hat{u}$ and $\hat{v}$ at their respective grid point locations.
\item Apply the initial and boundary conditions given in equations \eqref{e6} and \eqref{e7} respectively.
\item These equations will be solved algebraically because time advancement is fully explicit.
\item Linear stability conditions \eqref{e17}, \eqref{e18}, \eqref{e19}, and \eqref{e20} must be obeyed.
\item Divergence of the velocity field must be calculated at every time step, using the velocity field at the advanced time level, $(n +1)$,
\[
\nabla \cdot \vec{\mathbf{u}}= \dfrac{u_{i+1/2,j}^{n+1}-
u_{i-1/2,j}^{n+1}}{\Delta x}+ \dfrac{v_{i,j+1/2}^{n+1}-
v_{i,j-1/2}^{n+1}}{\Delta y}.
\]
\end{itemize}

The sum of the divergence magnitude at all grid points should be satisfied to machine zero at each time step. If this quantity increases, the calculation should be terminated and restarted with a smaller time step.

\subsubsection{Pressure Calculation}
\bigskip

\begin{itemize}
\item Calculate pressure from Pressure-Poisson equation \eqref{e21}.

\item Boundary condition applied to the pressure equation at all boundaries is the homogeneous Neumann boundary condition given by \eqref{e22}. This equation is solved at the pressure $(ij)$. Note that the Euler explicit time-advancement calculates the actual thermodynamic pressure (scaled by the constant density) and not a pseudo-pressure.
 \end{itemize}

\subsubsection{Velocity Correction}
\bigskip
\begin{itemize}
\item  Obtain the velocity field at the advanced time level $(n+1)$, for each velocity component, this equation gives
\begin{gather*}
u_{i+1/2,j}^{n+1}=\hat{u}_{i+1/2,j}-
\dfrac{\Delta t}{\Delta x}
\Big(P_{i+1,j}^{n+1}-P_{i,j}^{n+1}\Big),
\\[2mm]
v_{i,j+1/2}^{n+1}=\hat{v}_{i,j+1/2}-
\dfrac{\Delta t}{\Delta y}
 \Big(P_{i,j+1}^{n+1}-P_{i,j}^{n+1}\Big).
\end{gather*}
\end{itemize}

\subsubsection{Temperature Calculation}
\bigskip
\begin{itemize}
\item  Calculate the numerical solutions for temperature profiles by using equations \eqref{e26} and \eqref{e28}.
 \end{itemize}

\subsubsection{Concentration Calculation}
\medskip
\begin{itemize}
\item  Calculate the numerical solutions for concentration profiles by using equations \eqref{e27} and \eqref{e28}.
 \end{itemize}

\section{Results and Discussion}
\label{sec5.}

We used the modified Marker-And-Cell (MAC) method to carry out the numerical computations of the unknown flow variables $u$, $v$, $P$, $T$, and $C$ for the present problem. We have executed the modified MAC algorithm mentioned above with the aid of a computer program developed and run in C compiler. To verify our computer code, the numerical results obtained by the present method were compared with the benchmark results reported in \cite{b3}. It is seen that the results obtained in the present work are in good agreement with those reported in \cite{b3} at low Reynolds number $Re=100$. This indicates the validity of the numerical
code that we developed.

\begin{figure}[htb]
  \includegraphics[scale=0.68]{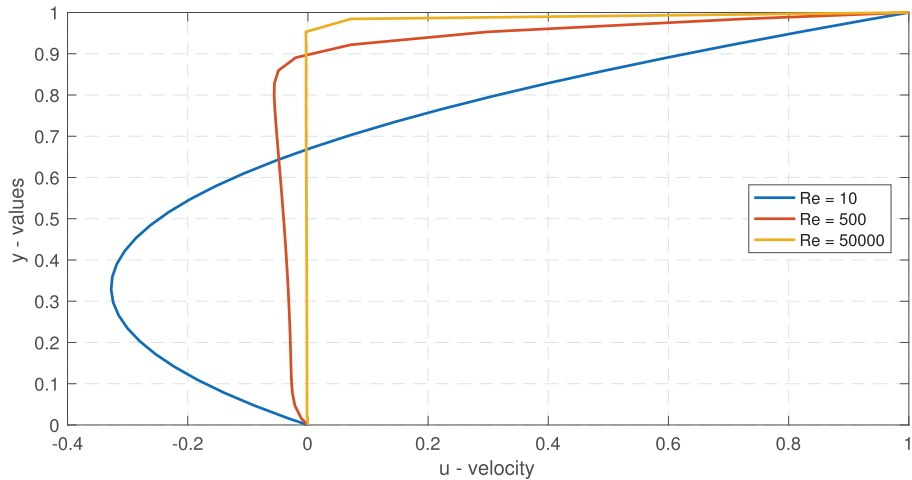}
  \caption{$u$-velocity along the vertical line through the geometric center of the domain at low, moderate, and high $Re$ for a fixed $Pr=6.62$ and $Sc=340$ for grid $32\times 32$ at time $t=0.0001s$.}
  \label{f4}
 \end{figure}

 Based on the numerical solutions for $u$-velocity, Figure \ref{f4} illustrates the variation of $u$-velocity along the vertical line through the geometric center of the rectangular domain at low, moderate, and high Reynolds numbers $Re=10$, $500$, and $50,000$. We can see that, for a given $Re=10$ and $Re=500$, $u$-velocity first decreases from the bottom boundary of the rectangular domain. It, then, increases to the upper boundary. But, for $Re=50,000$, $u$-velocity increases from the bottom boundary to the upper boundary of the rectangular domain. We also observe that, the absolute value of $u$-velocity decreases with increase in Reynolds number.

\begin{figure}[H]
  \includegraphics[scale=0.68]{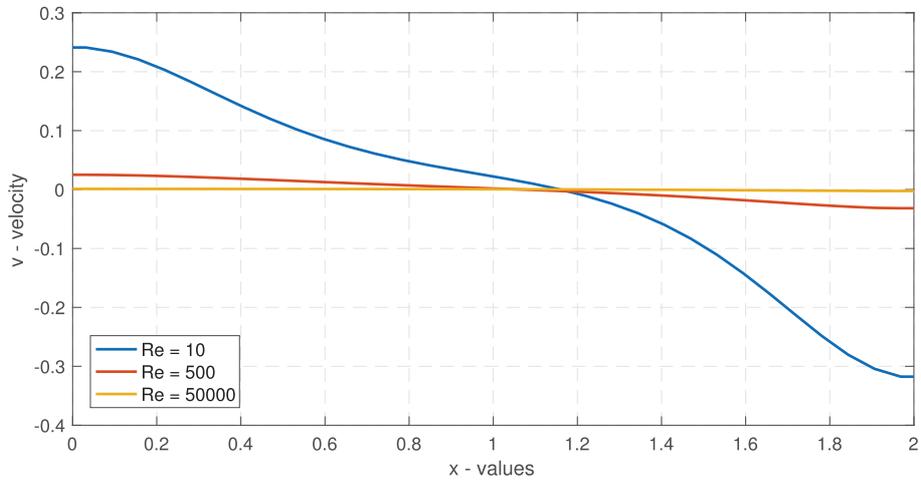}
  \caption{$v$-velocity along the horizontal line through the geometric center of the domain at low, moderate, and high $Re$ for a fixed $Pr=6.62$ and $Sc=340$ for grid $32\times 32$ at time $t=0.0001s$.}
  \label{f5}
 \end{figure}

 Based on the numerical solutions for $v$-velocity, Figure \ref{f5} illustrates the variation of $v$-velocity along the horizontal line through the geometric center of the rectangular domain. It is clear that for a given $Re$, $v$-velocity decreases from the left boundary to the right boundary. Further, the absolute value of $v$-velocity decreases with increase in Reynolds number.
 
\begin{figure}[H]
  \includegraphics[scale=0.68]{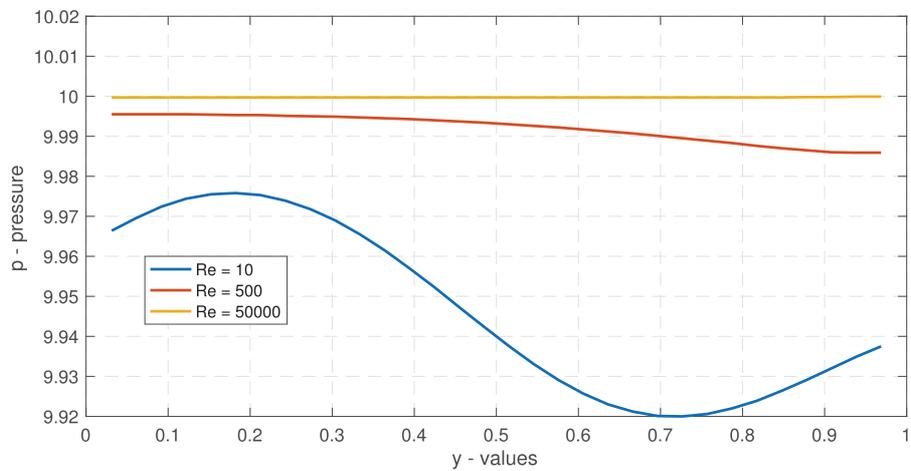}
  \caption{Pressure variation at low, moderate, and high $Re$ for a fixed $Pr=6.62$ and $Sc=340$ for grid $32\times 32$ at time $t=0.0001s$.}
  \label{f6}
 \end{figure}

Based on the numerical solutions for pressure, Figure \ref{f6} illustrates the variation of pressure in the rectangular domain. We observed that, for $Re=10$, the pressure is oscillatory in nature. However, for $Re=500$ and $Re=50,000$, we observed the pressure decrease from the left boundary to the right boundary. Further, the absolute value of pressure increases with increase in Reynolds number.

\begin{figure}[H]
  \includegraphics[scale=0.68]{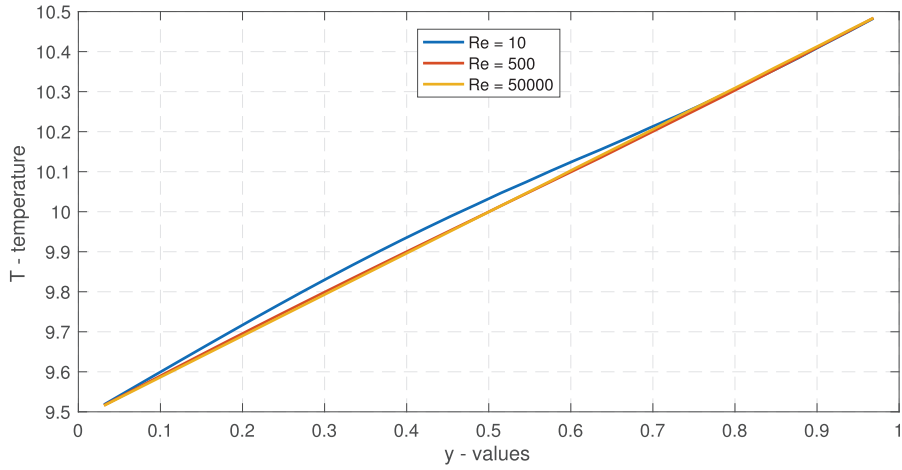}
  \caption{Temperature variation at low, moderate, and high $Re$ for a fixed $Pr=6.62$ and $Sc=340$ for grid $32\times 32$ at time $t=0.0001s$.}
  \label{f7}
 \end{figure}

Based on the numerical solutions for temperature, Figure \ref{f7} illustrates the variation  of temperature at different Reynolds numbers ($Re = 10$, $500$, and $50,000$), along the vertical line through the geometric center of the rectangular domain. It is clear that for a given $Re$, temperature increases from the bottom boundary to the upper boundary.

\begin{figure}[H]
  \includegraphics[scale=0.68]{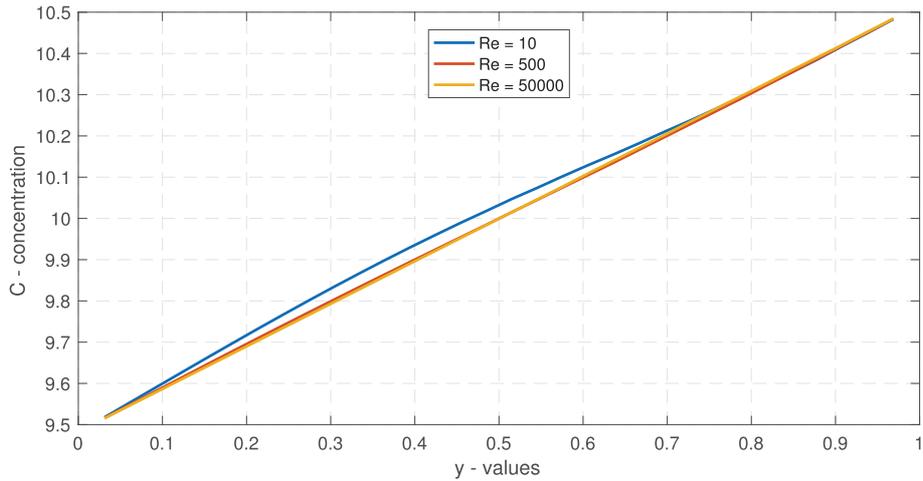}
  \caption{Concentration variation at low, moderate, and high $Re$ for a fixed $Pr=6.62$ and $Sc=340$ for grid $32\times 32$ at time $t=0.0001s$.}
  \label{f8}
 \end{figure}

 Based on the numerical solutions of concentration, Figure \ref{f8} illustrates the variation of concentration at different Reynolds numbers ($Re = 10$, $500$, and $50,000$), along the vertical line through the geometric center of the rectangular domain. It is clear that for a given $Re$, concentration increases from the bottom boundary to the upper boundary.

Grid dependence tests have been conducted on two grid systems
($16\times{16}$ and $32\times{32}$). In order to evaluate the effect  of the two grid systems on the natural convection flow in the rectangular domain, three representative quantities are evaluated numerically: the volumetric flow rate $(Q)$, average Nusselt number $(\overline{Nu})$, and average Sherwood number $(\overline{Sh})$ \cite{b26} across the horizontal centerline of the rectangular domain, which are defined as (also see \cite{b24, b25})
\begin{align}
&Q=\dfrac{1}{2}\int_{-\tfrac{L}{2}}^{\tfrac{L}{2}}|v|dx,
\\[1mm]
&\overline{Nu}=\dfrac{\dfrac{1}{L}
\int_{-\tfrac{L}{2}}^{\tfrac{L}{2}}\left| vT-\dfrac{\partial T}{\partial y} 
\right|dx}{\left (\dfrac{2\Delta{T}}{H} \right)},
\\[1mm]
&\overline{Sh}=\dfrac{\dfrac{1}{L}
 \int_{-\tfrac{L}{2}}^{\tfrac{L}{2}}\left| vC-\dfrac{\partial C}{\partial y} 
 \right |dx}{\left(
 \dfrac{2\Delta{C}}{H}\right)},
\end{align}
where local Nusselt number $(Nu)$, and local Sherwood number $(Sh)$ is defined as follows:
\begin{align}
&Nu=vT-\dfrac{\partial T}{\partial y},
\\[1mm]
&Sh=vC-\dfrac{\partial C}{\partial y}.
\end{align}
In the present problem length, $L$ of the  rectangular domain varies from 0 to 2, height, $H$ varies from $0$ to 1, $\Delta{T}=0.5$, and $\Delta{C}=0.5$. Then the volumetric flow rate $(Q)$, average Nusselt number $(\overline{Nu})$, and average Sherwood number $(\overline{Sh})$ across the horizontal centerline of the  rectangular domain, take the form
\begin{align}
&Q=\int_{0}^{2}|v|dx,
\\[1mm]
&\overline{Nu}=\int_{0}^{2}\left | vT-\dfrac{\partial T}{\partial y} \right |dx,
\\[1mm]
&\overline{Sh}=\int_{0}^{2}\left | vC-\dfrac{\partial C}{\partial y} \right |dx.
  \end{align}

 \begin{figure}[H]
  \includegraphics[scale=0.8]{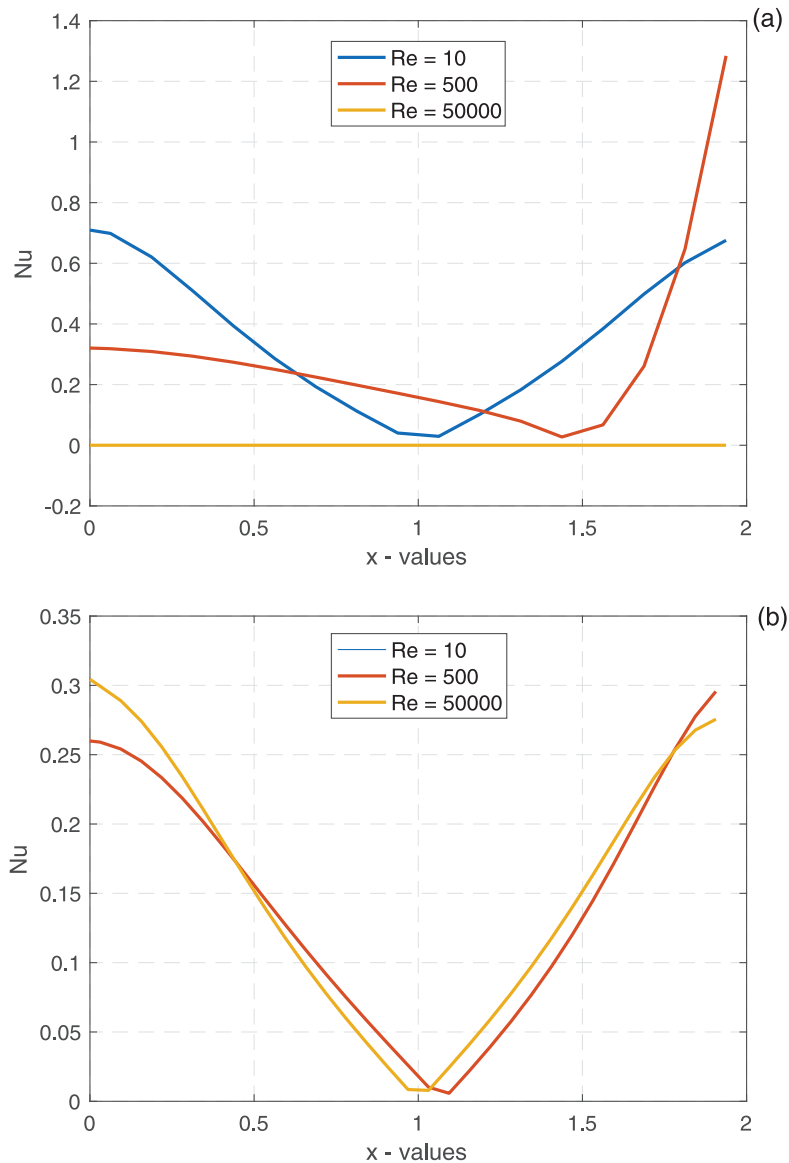}
  \caption{Variation of local Nusselt number ($Nu$) profiles along the horizontal line through the geometric center at low, moderate, and high $Re$ for a fixed $Pr=6.62$ and $Sc=340$, (a) for grid $16\times{16}$ and (b) for grid $32\times{32}$, at time $t=0.0001s$.}
  \label{f9}
 \end{figure}

 In order to investigate heat transfer from the bottom to the top wall of the rectangular domain, we have computed the numerical solutions of local Nusselt number for two different grid systems along the horizontal line through the geometric center of the domain at different Reynolds numbers. Figure \ref{f9} illustrates comparison of computed local Nusselt number ($Nu$) from the hot bottom wall to the cold top wall at time $t=0.0001s$. From this figure, we observe that, as we move along the horizontal line through the geometric center of the domain, heat transfer decreases up to the geometric center. It, then, increases symmetrically.

  \begin{figure}[H]
   \includegraphics[scale=0.8]{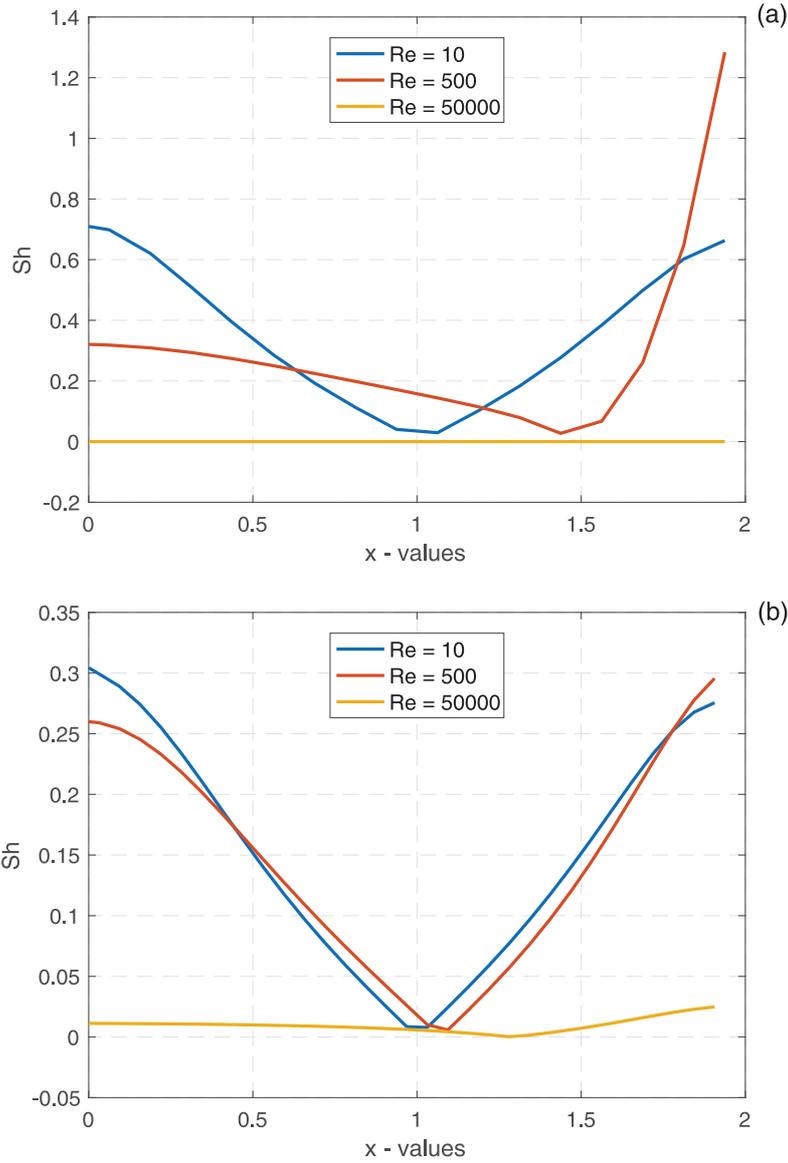}
   \caption{Variation of local Sherwood number ($Sh$) profiles along the horizontal line through the geometric center at low, moderate, and high $Re$ for a fixed $Pr=6.62$ and $Sc=340$, (a) for grid $16\times{16}$ and (b) for grid $32\times{32}$, at time $t=0.0001s$.}
   \label{f10}
  \end{figure}

 In order to investigate mass transfer from the bottom to the top wall of the rectangular domain, we have computed the numerical solutions of local Sherwood number for two different grid systems along the horizontal line through the geometric center of the domain at different Reynolds numbers. Figure \ref{f10} illustrates comparison of computed local Sherwood number ($Sh$) from the hot bottom wall to the cold top wall at time $t=0.0001s$. From this figure, we observe that, as we move along the horizontal line through geometric center of the domain, mass transfer decreases upto the geometric center. It, then, increases symmetrically.

 \section{Conclusions}
 \label{sec6.}

In this study, we have proposed a modified Marker-And-Cell (MAC) method to investigate the problem of an unsteady 2-D incompressible flow with heat and mass transfer at low, moderate, and high Reynolds numbers with no-slip and slip boundary conditions. We have used this method to solve the governing equations along with the boundary conditions and thereby to compute the flow variables, viz. $u$-velocity, $v$-velocity, $P$, $T$, and $C$. We have used the staggered grid approach of this method to discretize the governing equations of the problem. A modified MAC algorithm was proposed and used to compute the numerical solutions of the flow variables for Reynolds numbers $Re=10$, $500$, and $50,000$ in consonance with low, moderate, and high Reynolds numbers.

 Numerical solutions for $u$-velocity illustrates the variation of $u$-velocity along the vertical line through the geometric center of the rectangular domain at low, moderate, and high Reynolds numbers $Re=10$, $500$, and $50,000$. We have observed that, for a given $Re=10$ and $Re=500$, $u$-velocity first decreases from the bottom boundary of the rectangular domain. It, then, increases to the upper boundary. But, for $Re=50,000$, $u$-velocity increases from the bottom boundary to the upper boundary of the rectangular domain. We also observed that the absolute value of $u$-velocity decreases with increase in Reynolds number. The numerical solutions for $v$-velocity illustrates the variation of $v$-velocity along the horizontal line through the geometric center of the rectangular domain. We have observed that, for a given $Re$, $v$-velocity decreases from the left boundary to the right boundary. Further,we also observed that the absolute value of $v$-velocity decreases with increase in Reynolds number.

 Numerical solutions for pressure illustrates the variation of pressure in the rectangular domain. We have observed that, for $Re=10$, the pressure is oscillatory in nature. However, for $Re=500$ and $Re=50,000$, we observed the pressure decrease from the left boundary to the right boundary. Further, we have  also observed that,the absolute value of pressure increases with increase in Reynolds number. The numerical solutions for temperature illustrates the variation  of temperature at different Reynolds numbers ($Re = 10$, $500$, and $50,000$), along the vertical line through the geometric center of the rectangular domain. We have observed that, for a given $Re$, temperature increases from the bottom boundary to the upper boundary. The numerical solutions of concentration, illustrates the variation of concentration at different Reynolds numbers ($Re = 10$, $500$, and $50,000$), along the vertical line through the geometric center of the rectangular domain. We have observed that, for a given $Re$, concentration increases from the bottom boundary to the upper boundary.

  Based on the computed local Nusselt number($Nu$) from the hot bottom wall to the cold top wall at time $t=0.0001s$, we have observed that, as we move along the horizontal line through the geometric center of the domain, heat transfer decreases up to the geometric center. It, then, increases symmetrically. Based on the computed local Sherwood number ($Sh$) from the hot bottom wall to the cold top wall at time $t=0.0001s$.we have observed that, as we move along the horizontal line through geometric center of the domain, mass transfer decreases upto the geometric center. It, then, increases symmetrically.

 \section{Code Validation}
 \label{sec7.}
 To check the validity of our present computer code used to obtain the numerical
 results of $u$-velocity and $v$-velocity, we have compared our present results with
 those benchmark results are given by Ghia et al. \cite{b3} and it has been found that
 they are in good agreement.

\begin{figure}[H]
   \includegraphics[scale=0.7]{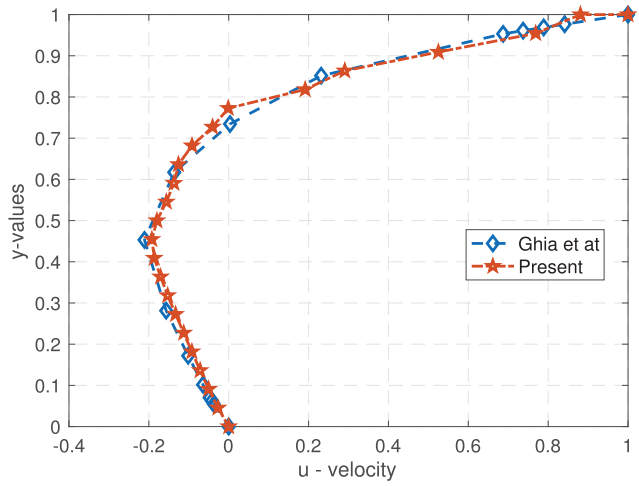}
   \caption{Comparison of the numerical results of u-velocity along
    the vertical line through the geometric center of the square cavity for $Re=100$.}
   \label{f11}
  \end{figure}

   \begin{figure}[H]
    \includegraphics[scale=0.7]{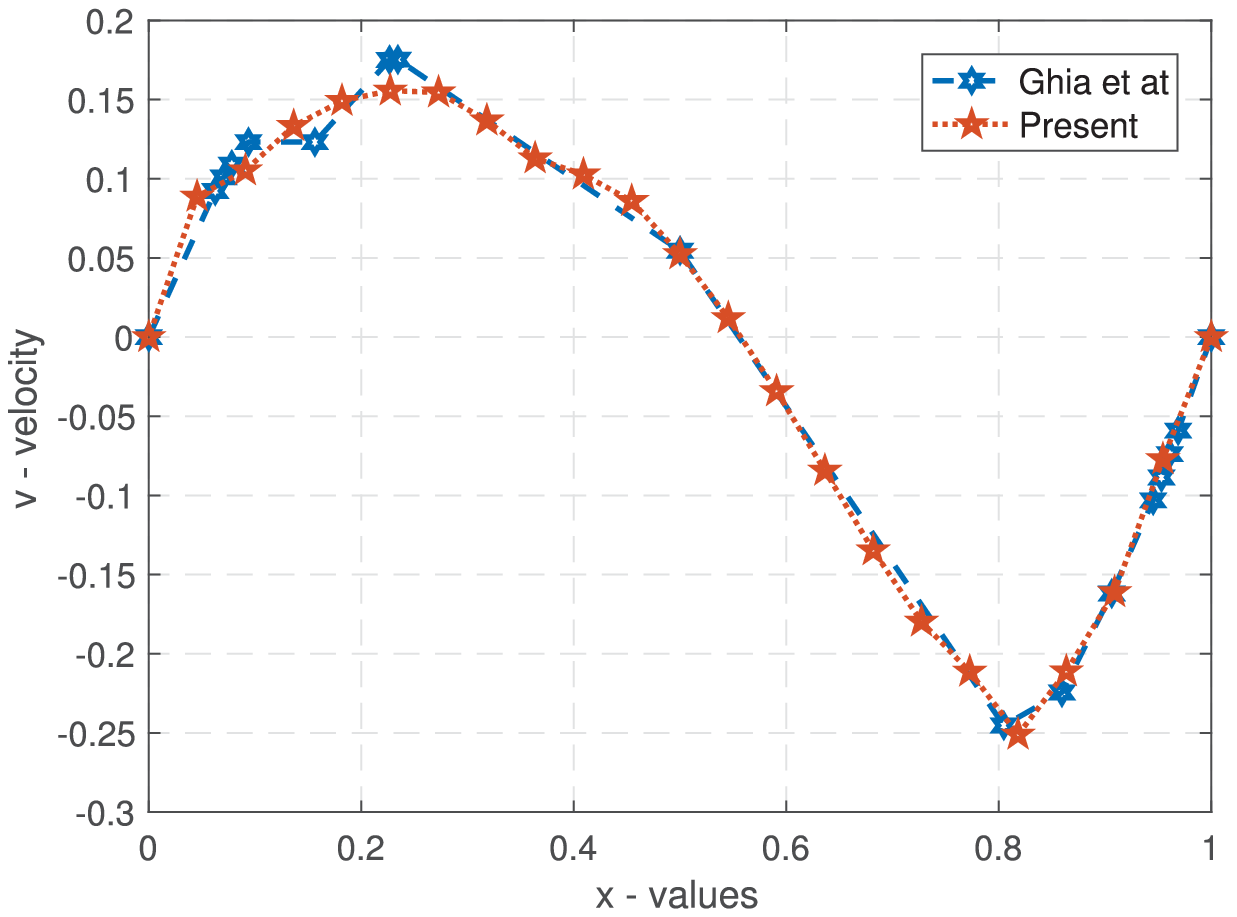}
    \caption{Comparison of the numerical results of v-velocity along
     the horizontal line through the geometric center of the square cavity for
     $Re=100$.}
    \label{f12}
   \end{figure}
\mbox{}

\nomenclature{$p$}{pressure (${\text{N}/\text{m}^2}$)}
\nomenclature{$P$}{dimensionless pressure $\left(\dfrac{p}{\rho}\right)$ that has been with respect to $\rho u^2$}
\nomenclature{$i$}{index used in tensor notation}
\nomenclature{$j$}{index used in tensor notation}
\nomenclature{\text{Re}}{Reynolds number, $\mbox{uL}/\mathrm{\mu}$}
\nomenclature{\text{Sc}}{Schmidt number, $\mathrm{\nu} /\mbox{D}$}
\nomenclature{\text{Pl}}{Prandtl number, $\mathrm{\nu}/\mbox{k}$}
\nomenclature{\text{Nu}}{local Nusselt number}
\nomenclature{\text{Sh}}{local Sherwood number}
\nomenclature{$\overline{\text{Nu}}$}{average Nusselt number}
\nomenclature{$\overline{\text{Sh}}$}{average Sherwood number}
\nomenclature{$Q$}{volumetric flow rate ($\text{m}^3/\text{s}$)}
\nomenclature{$k$}{thermal diffusivity ($\text{m}^2/\text{s}$)}
\nomenclature{$D$}{mass diffusivity ($\text{m}^2/\text{s}$)}
\nomenclature{$\mu$}{viscosity of the fluid ($\text{N}\cdot \text{m}^2/\text {s}$)}
\nomenclature{$\nu$}{kinematic viscosity ($\text{m}^2/\text{s}$)}
\nomenclature{$T$}{temperature ($\text{K}$)}
\nomenclature{$C$}{concentration ($\text{kg}/\text{m}^3$)}
\nomenclature{$\vec{\mathbf{u}}$}{ velocity vector}
\nomenclature{$t$}{non-dimensional time ($\text{s}$)}
\nomenclature{$x$, $y$}{coordinates}
\nomenclature{$\Delta x$}{grid spacing along $x$-axis}
\nomenclature{$\Delta y$}{grid spacing along $y$-axis}
\nomenclature{$\Delta t$}{time spacing}
\nomenclature{$\nabla \cdot \vec{\mathbf{u}}$}{divergence of velocity vector}
\nomenclature{$\nabla \cdot \hat{u}$}{divergence of pseudo-velocity vector}
\nomenclature{$u$, $v$}{velocity components in $x$, $y$ coordinate directions, respectively ($\text{m}/\text{s}$)}
\nomenclature{$\hat{u}$, $\hat{v}$}{pseudo-velocity components in $x$, $y$ coordinate directions, respectively ($\text{m}/\text{s}$)}
\nomenclature{$u^n$}{$x$ component of the velocity after $n$ iterations}
\nomenclature{$v^n$}{$y$ component of the velocity after $n$ iterations}
\nomenclature{$\dfrac{\partial}{\partial n}$}{differentiation along the normal to the boundary}
\nomenclature{$t^n$}{time level after $n$ iterations}
\nomenclature{$\rho$}{fluid density ($\text{kg}/\text{m}^3$)}
\nomenclature{$n$}{pertains to the $n$ iteration}

\printnomenclature

 \section*{Acknowledgments}

 The first author acknowledge the support from research council, University of Delhi for providing research and development grand 2015--16 vide letter No. RC/2015-9677 to carry out this work.


\end{document}